# Tuning Structural, Transport and Magnetic Properties of Epitaxial SrRuO$_3$ through Ba-Substitution

Zeeshan Ali[1*], Zhen Wang[1,2,3], Alessandro R. Mazza[4,5], Mohammad Saghayezhian[1], Roshan Nepal[1], Thomas Z. Ward[4], Yimei Zhu[3,†], and Jiandi Zhang[1,6,‡]

[1]*Department of Physics & Astronomy, Louisiana State University, Baton Rouge, LA 70803, USA.*

[2]*University of Science and Technology of China, Hefei, Anhui 230026, People's Republic of China.*

[3]*Condensed Matter Physics & Materials Science, Department, Brookhaven National Laboratory, Upton, NY 11973, USA.*

[4]*Materials Science and Technology Division, Oak Ridge National Laboratory, Oak Ridge, Tennessee 37831, USA.*

[5]*Center for Integrated Nanotechnologies, Los Alamos National Laboratory, Los Alamos, New Mexico 87545, USA.*

[6]*Beijing National Laboratory for Condensed Matter Physics, Institute of Physics, Chinese Academy of Sciences, Beijing 100190, People's Republic of China.*

The perovskite ruthenates ($A$RuO$_3$, $A$ = Ca, Ba, or Sr) exhibit unique properties owing to a subtle interplay of crystal structure and electronic-spin degrees of freedom. Here, we demonstrate an intriguing continuous tuning of crystal symmetry from orthorhombic to tetragonal (no octahedral rotations) phases in epitaxial SrRuO$_3$ achieved via Ba-substitution (Sr$_{1-x}$Ba$_x$RuO$_3$ with $0 \leq x \leq 0.7$). An initial Ba-substitution to SrRuO$_3$ not only changes the ferromagnetic properties, but also tunes the perpendicular magnetic anisotropy via flattening the Ru-O-Ru bond angle (to 180°), resulting in the maximum Curie temperature and an extinction of RuO$_6$ rotational distortions at $x \approx 0.20$. For $x \leq 0.2$, the suppression of RuO$_6$ octahedral rotational distortion dominantly enhances the ferromagnetism in the system, though competing with the impact of the RuO$_6$ tetragonal distortion. Further increasing $x > 0.2$ gradually enhances the tetragonal-type distortion, resulting in the tuning of Ru-4$d$ orbital occupancy and suppression of ferromagnetism. Our results demonstrate that isovalent substitution of the $A$-site cations significantly and controllably impacts both electronic and magnetic properties of perovskite oxides.

*zee89ali@gmail.com
†zhu@bnl.gov
‡jiandiz@iphy.ac.cn

## I. INTRODUCTION

Perovskite oxides ($ABO_3$) show a wide range of emergent functionalities originating from electro-magnetic interactions coupled to the octahedral units [1,2]. In a bulk perovskite oxide system, the oxygen octahedral environment is generally controlled via conventional chemical substitution or external stimuli such as hydrostatic pressure or temperature [3–10]. On the other hand, artificial heterostructures can provide a platform with additional structural tuning routes such as epitaxial-strain [11–16], interfacial-octahedral engineering [1,17–20], strain-controlled doping etc. [21–23]. Heterostructure engineering has been intensively explored in perovskites, due to the strong structure-property coupling which can drive new ground states that are not accessible in bulk [24–29]. In particular, it has been shown that imposing an artificial heterostructure geometry allows effective tuning of octahedral bond angle, which can be used to manipulate spin alignment and magnetization dynamics [17,18,30–35].

Perovskite ruthenates ($ARuO_3$ $A$= Ca, Sr, or Ba) are an excellent candidate to explore the connection of magnetic and electronic properties with the local control of structure. In this system, isovalent substitution has been shown to drastically affect the physical properties due to the close coupling between structural and electron-spin degrees of freedom [36–38]. The replacement of $Sr^{2+}$ (1.44 Å) ions with $Ca^{2+}$ (1.34 Å) ions in bulk-family preserves orthorhombic symmetry, yet distinct Ru-O-Ru angles ($CaRuO_3$:148° and $SrRuO_3$:163°) trigger a subtle modification in the Fermi-level density of states, which leads to a non-magnetic $CaRuO_3$ (CRO) and ferromagnetic $SrRuO_3$ (SRO) [36–40]. In contrast, the larger $A$-site cation $Ba^{2+}$ (1.61 Å) introduces a transition from orthorhombic to cubic-symmetry and a suppression of ferromagnetism [from $T_c \approx 160$ K in SRO to 60 K in $BaRuO_3$] [38]. Since the nominal electron counting is not changing, the octahedral angle, as well as $A$-O and Ru-O bonding nature appear to be the controlling factor, but the dominant driver of this change has still not been identified. How the electronic structure and magnetism evolve with different substitution levels are far from clear. A complicating factor has been the difficulty in synthesizing the Ba-doped SrRuO$_3$ single crystals, where high pressures (18 GPa) are necessary to stabilize a perovskite structure [37,38]. This makes engineering perovskite functionalities dependent on finding innovative synthesis routes. In this respect, the application of substrate-induced epitaxial strain delivers an auxiliary approach to high-pressure bulk synthesis and provides a viable platform to stabilize crystalline films with novel phases and multi-functionalities [11,41–45].

In this study, we demonstrate epitaxial stabilization of the Sr$_{1-x}$Ba$_x$RuO$_3$ ($0 \leq x \leq 0.7$) thin films using strain engineering. Combining atomically resolved scanning transmission electron microscopy (STEM) imaging, electron energy-loss spectroscopy (EELS), and X-ray diffraction (XRD) along with magneto-transport measurements, we reveal that the lattice structure can be continuously tuned in a series of Sr$_{1-x}$Ba$_x$RuO$_3$ ($0 \leq x \leq 0.7$) thin films grown on SrTiO$_3$ (001). Ba-cation substitution ($x$) is found to drive the structure from bulk-like orthorhombic ($x = 0$) to a tetragonal (no octahedron rotations: $a^0b^0c^0$) phase ($x = 0.2$) enabled by change of RuO$_6$ distortions. Doping with Ba not only transforms the lattice symmetry but also triggers a modification in Curie temperature ($T_c$) and perpendicular magnetic anisotropy (PMA). The resultant tetragonal (without octahedron rotation) structured film at $x \approx 0.2$ is found to be ferromagnetic with the $T_c \approx 145$ K and exhibits the strongest PMA. However, further increasing Ba-substitution considerably suppresses the ferromagnetism, especially as $x \geq 0.5$. These results demonstrate that such isovalent $A$-site substitution has a significant impact on lattice structure which can be used to manipulate electronic and magnetic functionalities.

## II. EXPERIMENTAL METHODS

**Film targets:** The series of $Sr_{1-x}Ba_xRuO_3$ ($x$ = 0, 0.08, 0.2, 0.5, 0.7) thin film targets were synthesized by conventional solid-state reaction. The starting materials of $SrCO_3$, $BaCO_3$, and $RuO_2$ in stoichiometric ratios were first mixed thoroughly and then heated at 1200° C in the air for 48 h. Regrinding and sintering at 1200 ° C were performed to increase the chemical homogeneity. The resultant powders were grounded and pressed into pellets (one-inch diameter) under a hydraulic pressure of 1000 psi. The target pellets were then sintered at 1100° C for 48 h in an oxygen atmosphere. To increase the chemical homogeneity and target density, another sintering at 1100°C for 48 h in an oxygen atmosphere was performed.

**Films growth:** The thin films of $Sr_{1-x}Ba_xRuO_3$ were grown by pulsed laser deposition (PLD) at 700° C with an oxygen pressure of 100 mTorr on $TiO_2$ terminated $SrTiO_3$ (001) (STO) substrates. A KrF excimer laser ($\lambda$ = 248 nm) with a 10 Hz repetition rate, and with an energy of 300 mJ (laser energy density ~ 1 J cm$^{-2}$) was focused on $Sr_{1-x}Ba_xRuO_3$ targets. Post deposition, the films were cooled down at ~10°/min to room temperature in an oxygen atmosphere of 100 mTorr. To monitor the film growth, the *in-situ* differentially-pumped reflective high energy electron diffraction (RHEED) was employed. The film thickness was kept at ~ 40-unit cells (u.c.).

**Scanning transmission electron microscopy and electron energy loss spectroscopy:** STEM and EELS experiments were performed on a 200 kV JEOL ARM electron microscope at Brookhaven National Library equipped with double aberration correctors, a dual energy-loss spectrometer, and a cold field-emission source. TEM samples were prepared using a focused ion beam with Ga$^+$ ions followed by Ar$^+$ ions milling to a thickness of ~30 nm. The atomic-resolution STEM images were collected with a 21 mrad convergent angle (30 μm condenser aperture) and a collection angle of 67 – 275 mrad for high-angle annular dark-field (HAADF) and 11 – 23 mrad for annular bright-field (ABF) imaging. The atomic positions were obtained using two-dimensional Gaussian fitting following the maximum intensity. The microscope conditions were optimized for EELS acquisition with a probe size of 0.8 Å, a convergence semi-angle of 20 mrad, and a collection semi-angle of 88 mrad. Dual EELS mode was used to collect low-loss and core-loss spectra simultaneously for energy drift calibration in the collecting process. EELS mapping was obtained across the whole film with a step size of 0.2 Å and a dwell time of 0.05 s/pixel. EELS background was subtracted using a power-law function, and multiple scattering was removed by a Fourier deconvolution method.

**X-ray diffraction:** Panalytical X'Pert thin-film diffractometer with Cu Kα−1 radiation and a single-crystal monochromator was employed for coupled scans, omega scans, and reciprocal space mapping of XRD. The half-order integer spot measurements were carried out in a lab-based Malvern Panalytical X'Pert four-circle diffractometer with a collection time of 100-250 sec/point.

**Electron and Magneto-transport:** The magnetization was measured by using a Quantum Design Superconducting Quantum Interference Device, a reciprocating sample option. The electron transport measurements were performed in a four-probe configuration on a Quantum Design Physical Property measurement system.

## III. RESULTS AND DISCUSSION

### A. Crystal Structure

We have fabricated a series of $Sr_{1-x}Ba_xRuO_3$ ($0 \leq x \leq 0.7$) thin-films on $SrTiO_3$ (001) substrates with a thickness of 40 unit-cells (u.c.). **FIG. 1(a-b)** shows the X-ray diffraction (θ-2θ) coupled scan for $Sr_{1-x}Ba_xRuO_3$ thin films. In **FIG. 1(a-b)**, the $SrTiO_3$ $(002)_{pc}$ Bragg's reflection (*) and film peaks (marked with arrows) as well as Laue interferences could be observed, confirming epitaxial stability and good crystallinity of thin films. The XRD of the heteroepitaxial films demonstrates that the perovskite-structure can be stabilized using pulsed laser deposition. These results demonstrate the effectiveness of strain-induced epitaxial stabilization of pure-phase perovskite films. This is an important observation as the $Sr_{1-x}Ba_xRuO_3$ perovskite-structure in bulk can only be stabilized using high pressure synthesis [38]. The ambient pressure synthesis favors the development of different polymorphs such as the nine layered rhombohedral (9R), six layered

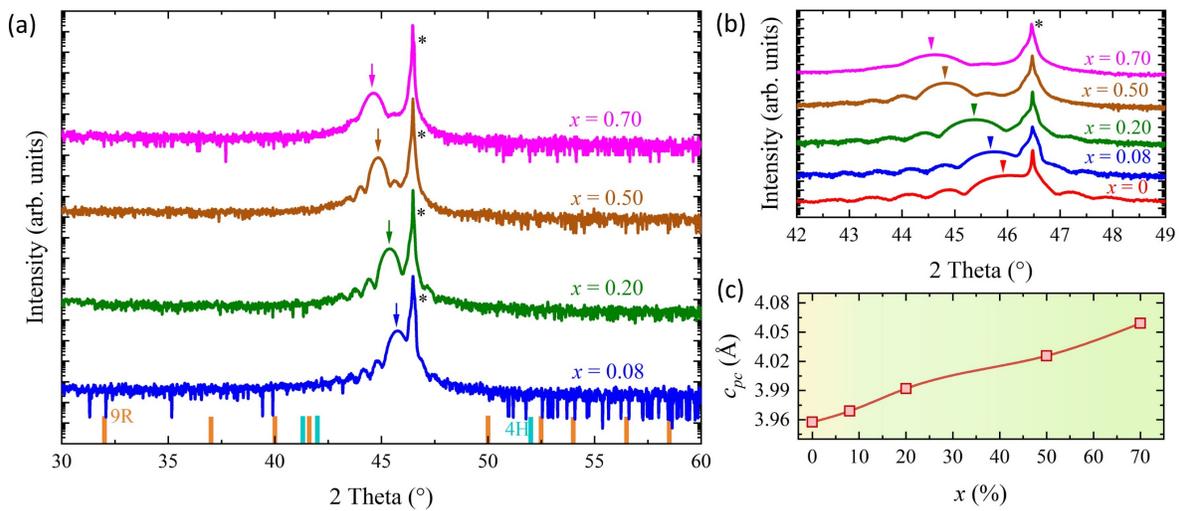

**FIG. 1.** XRD characterization of $Sr_{1-x}Ba_xRuO_3$ ($0 \leq x \leq 0.7$) thin films at room temperature: (a) Long-range coupled ($\theta$-$2\theta$) scan, (b) $\theta$-$2\theta$ scan around (002) spot and (c) Obtained out-of-plane pseudocubic ($c_{pc}$) lattice constant versus Ba-concentration ($x$). The $SrTiO_3$ $(002)_{pc}$ Bragg's reflection and film peaks are marked with * and arrows, respectively.

hexagonal (6H), and four layered hexagonal structures (4H) [37,38,46].

The lattice constant extracted from the specular XRD illustrates that the out-of-plane (OP) lattice parameter ($c_{pc}$) could be systematically modified via Ba substitution in the presence of epitaxial strain [**FIG. 1(c)**]. For the $SrRuO_3$ film, the OP lattice parameter is $c_{pc}$ = 3.957 Å, where elongation in the OP direction is caused by compressive strain (bulk $c_{pc,\,STO}$ = 3.905 Å < $c_{pc,\,SRO}$ =

3.925 Å). The systematic Ba-doping of SRO results in $(002)_{pc}$ peak shift to lower $2\theta$-angles [**FIG. 1(a-b)**]. The shifting of the peak demonstrates that the OP lattice parameter expands and is summarized as a function of Ba-doping concentration ($x$) in **FIG. 1(c)**. Such OP lattice enlargement with varying degrees of Ba cation substitution advocates a structural transition.

To further understand the crystal structure of $Sr_{1-x}Ba_xRuO_3$ thin films, we performed reciprocal space mappings (RSM) around $(103)_{pc}$ SrTiO$_3$ Bragg reflections. The RSM of a SrRuO$_3$ film is shown in **FIG. 2(a)**. The inspection of film peaks illustrates different $Q_z$ values for $(103)_{pc}$, $(013)_{pc}$, $(-103)_{pc}$ and $(0-13)_{pc}$ diffraction peaks, suggesting an orthorhombic structure. The observation agrees with a GdFeO$_3$-type orthorhombic structure [15,47,48]. Furthermore, the film diffraction peaks have identical $Q_x$ with STO as presented in **FIG. 2(a)**, confirming the SRO film is coherently strained. In contrast, for the $Sr_{0.92}Ba_{0.08}RuO_3$ [**FIG. 2(b)**], the film diffraction peaks have matching $Q_z$, indicating a tetragonal structural variant. Increasing Ba-substitution to 20%, preserves uniform $Q_z$ distribution while maintaining epitaxy to STO but with larger OP expansion [**FIG. 2(c)**].

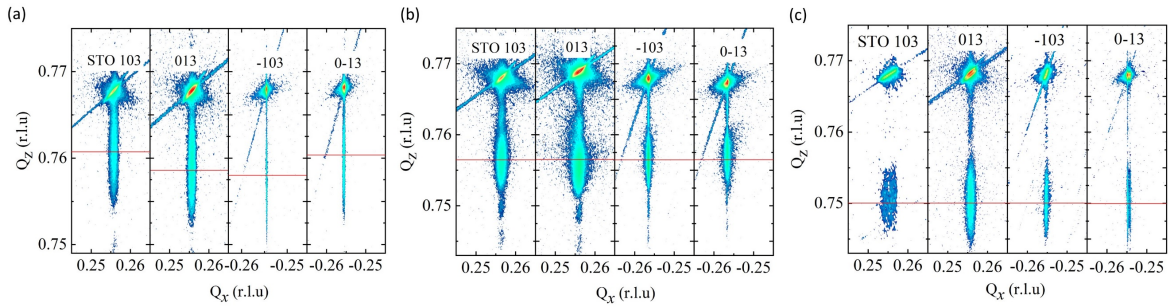

**FIG. 2.** XRD reciprocal space mapping about the $(103)_{pc}$ reflections from (a) $x = 0$, (b) $x = 0.08$, and (c) $x = 0.2$ of $Sr_{1-x}Ba_xRuO_3$ thin films.

**B.     Atomic-scale structure and composition**

We have performed STEM and EELS to reveal microscopically the atomic structure and chemical composition of the films [**FIG. 3**]. The high-angle-annular-dark-field (HAADF)-STEM (*Z*-contrast) images show coherent growth of high-quality $Sr_{0.92}Ba_{0.08}RuO_3$ [**FIG. 3(a-b)**] and $Sr_{0.8}Ba_{0.2}RuO_3$ [**FIG. 3(d-e)**] films on the STO substrate. The large area HAADF and highly magnified HAADF corroborate dislocation-free structures possessing coherent interfaces and high quality. EELS is also used to extract the elemental profiles of Ba and Sr elements as given in **FIG. 3(c)** and **FIG. 3(f).** The qualitative elemental profiles were obtained by integrating EELS intensity maps from different regions. The EELS data confirms a homogeneous distribution of doped Ba in the films. The averaged Ba concentration was measured to be $0.07 \pm 0.004$ in the $Sr_{0.92}Ba_{0.08}RuO_3$ film and 0.20 ± 0.017 in the $Sr_{0.8}Ba_{0.2}RuO_3$ film, which supports the desired Ba-doping concentration of 8 % and 20 %, respectively.

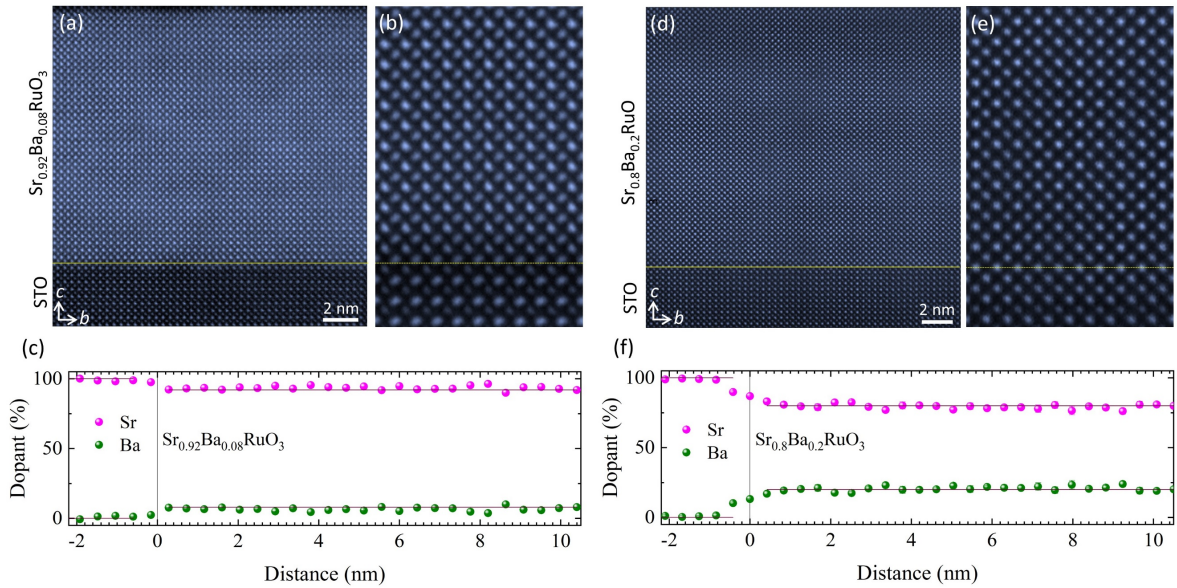

**FIG. 3.** (a) Large area high-angle annular dark-field (HAADF)-STEM, and (b) highly magnified HAADF images taken along the [100]-STO direction for $Sr_{0.92}Ba_{0.08}RuO_3$ film. The yellow line marks the substrate-film interface. (c) Dopant-concentration (Sr/Ba) profile of Sr (pink) and Ba (green) extracted from Sr-L and Ba-L edges for $Sr_{0.92}Ba_{0.08}RuO_3$. (d) Large area, (e) high magnification HAADF-STEM images taken along the [100]-STO direction, and Dopant-concentration (Sr/Ba) profile of Sr (pink) and Ba (green) for $Sr_{0.8}Ba_{0.2}RuO_3$ film.

## C.    $RuO_6$ octahedra rotations

To shed light on structural transitions with Ba-doping, we have investigated the $RuO_6$ octahedra rotations of $Sr_{1-x}Ba_xRuO_3$ thin films using a combination of atomically resolved scanning/transmission electron microscopy (S/TEM) and half-integer X-ray reflections. In $ABO_3$ perovskites, the Glazer notation is used to describe octahedral rotations [49,50]. In this picture, the $BO_6$ octahedra rotates about the three orthogonal pseudo-cubic (*pc*) crystallographic axes: $[100]_{pc}$, $[010]_{pc}$, and $[001]_{pc}$. The $BO_6$ rotational magnitudes are specified by letters *a*, *b*, and *c*, referring to the $[100]_{pc}$, $[010]_{pc}$, and $[001]_{pc}$ axes, respectively. The superscripts +, –, or 0 are used to signify whether the adjacent octahedra around one axis rotates in-phase (+), out-of-phase (–), or does not rotate (0). In X-ray diffraction, the in-phase rotations (+) give rise to ½ (odd-odd-even) type of reflections, while out-of-phase (–) produce ½ (odd-odd-odd) reflections [48,51–53].

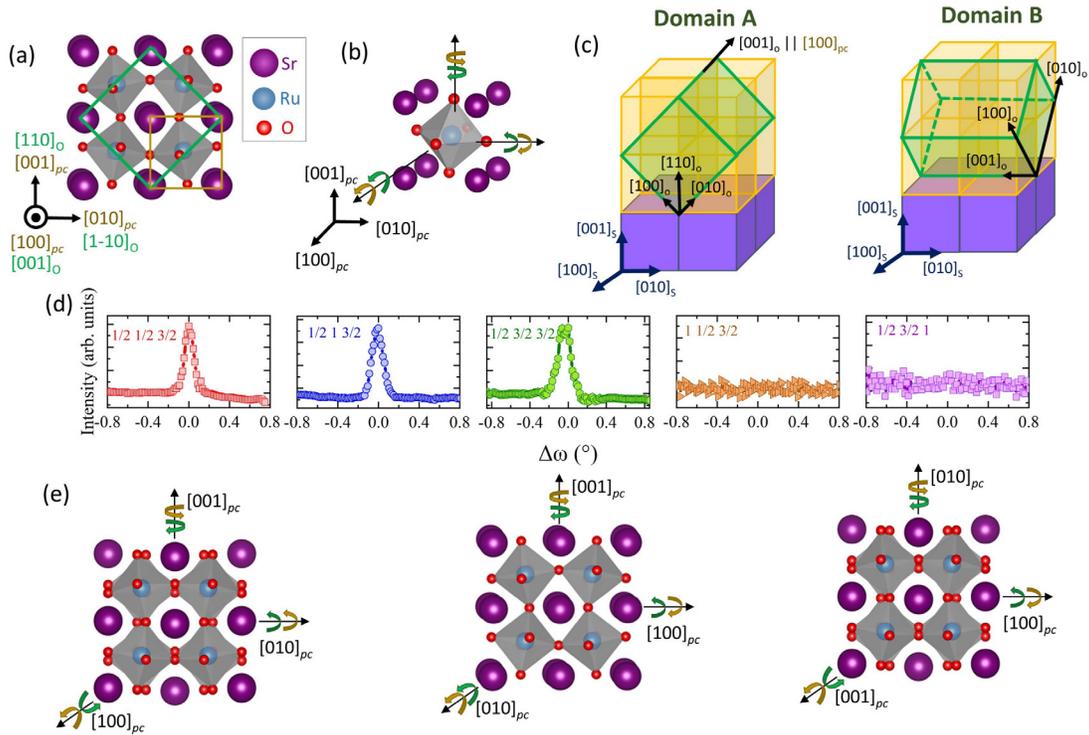

**FIG. 4.** (a) Crystal structure of bulk $SrRuO_3$ (SRO). The orthorhombic (O) and pseudo-cubic (pc) unit cells are indicated by green and orange color, respectively. (b) Pseudocubic unit-cell of SRO. The octahedral rotation is in-phase along the $[100]_{pc}$ ($[001]_O$), out-of-phase about the $[010]_{pc}$ ($[1\text{-}10]_O$), and $[001]_{pc}$ ($[110]_O$) axes, respectively. (c) Schematic illustrations of orthorhombic SRO unit-cell orientation on a cubic $SrTiO_3$ substrate with two possible orientations (domain A and B). The in-phase rotation (+) axis $[100]_{pc}$ of the film lies within the film plane, which aligns along the $[100]_S$ of the substrate in Domain A, and along $[010]_S$ in Domain B. (d) Half-integer X-ray reflections for SRO. (e) Representation of octahedral rotation pattern in SRO film. Here, the coordinates are defined with respect to pseudocubic lattice.

Bulk SrRuO3 falls into the orthorhombic tilt category $a^+b^-c^-$ (*Pbnm* space group) as identified by the presence of an in-phase and two out-of-phase rotations [**FIG. 4(a-b)**]. Generally, SrRuO3 films on SrTiO3 substrates grow in the $[110]_O$ direction, whereas the subscript O stands for orthorhombic notation [14,54]. Thus, providing two possible orthorhombic unit-cell orientations, as given schematically in **FIG. 4(c)**. Consequently, either in-phase $[100]_{pc}$ or out-of-phase $[010]_{pc}$ axes align with the $[100]_S$ direction of the STO substrate, leading to two possible domain structures [**FIG. 4(c)**] referred to as A and B. Here, the subscripts stand for orthorhombic (O), pseudo-cubic (pc), and substrate (S). In domain A, the in-phase rotation (+) axis $[100]_{pc}$ of the film lies parallel to the $[100]_S$ of the substrate, i.e., $[100]_{pc}//[100]_S$ [**FIG. 4(c)**]. In domain B, the in-phase rotation (+) pseudo-cubic $[100]_{pc}$ film axis is aligned with the $[010]_S$ STO-axis, i.e., $[100]_{pc}//[010]_S$ [**FIG. 4(c)**]. The structure in domains A and B are equivalent, but rotated 90° relative to one another [48,54,55]. In our case, the half-angle reflections in **FIG. 4(d)** establish a single domain (i.e., domain B) structure with an $a^-b^+c^-$ rotation pattern for SrRuO3 film. In **FIG. 4(d)**, the $\left(\frac{1}{2},\frac{1}{2},\frac{3}{2}\right)$ peak reflects $a^-$ rotation, while the absence of the $\left(1,\frac{1}{2},\frac{3}{2}\right)$ excludes the possibility of $a^+$. Along the $[010]_{pc}$ the $\left(\frac{1}{2},1,\frac{3}{2}\right)$ peak suggests $b^+$ octahedral rotations. Lastly, the occurrence of $\left(\frac{1}{2},\frac{3}{2},\frac{3}{2}\right)$ advocates $c^-$, and the absence of the $\left(\frac{1}{2},\frac{3}{2},1\right)$ rules out $c^+$ rotation. From these results; the $a^-b^+c^-$ (or equivalently $a^+b^-c^-$ as $a = b$) rotation pattern is valid for SrRuO₃ film, consistent with that of the bulk [21,51]. The preference for Domain B-type structures could be attributed to SrTiO₃ substrate vicinality since the modified miscut angle and topography of substrate step edges tends to determine the alignment of the orthorhombic $c$-axis [41,45]. Previous studies have reported that on exactly 001-oriented STO substrate, the volume fraction for A and B-type domains could be identical, though, depending on the vicinal nature of the (001)-SrTiO₃ substrate, one kind of domain formation is preferred [56,57].

In compliment to this XRD result, high-resolution STEM images and corresponding Fast Fourier Transform (FFT) patterns [**FIG. 5**] of $Sr_{0.92}Ba_{0.08}RuO_3$ film shows that the majority of regions (~80%) are Domain B ($[010]_{pc}//[100]_S$). The coexistence of domain A/B, suggests that in-plane rotations are either − or +, respectively. Due to the coexisting domain A, and B structure [see **FIG. 5(a-b)**], we can acquire simultaneously the projected structures in the $[100]_{pc}$ and $[010]_{pc}$ directions when viewing along the $[100]_S$ direction of the substrate. **FIG. 5(c-d)** depicts HRTEM images and corresponding FFT patterns, whereas the experimental FFT's are in accordance with the simulated diffraction patterns in the $[100]_{pc}$ [Domain A; see **FIG. 5(a)** and **FIG. 5(c)**] and $[010]_{pc}$ [Domain B: **FIG. 5(b)** and **FIG. 5(d)**] directions. The two-domain structures could be distinguished by the presence of characteristic diffraction spots marked as green and red in **FIG. 5(c-d)**. The structures of domains A and B are identical, but when along the $[001]_S$ direction, as in **FIG. 5(a-b)**, we can observe simultaneously the RuO₆ tilt in Domain A, and $A$-site cation displacement in Domain B [54]. To shed light on octahedra distortion, we have acquired atomically resolved HAADF STEM images of the $Sr_{0.92}Ba_{0.08}RuO_3$ film [**FIG. 5(e-g)**]. The STEM images validate an $A$-site cation displacement forming a zig-zag pattern [**FIG. 5(e-g)**]. The quantitative analysis of *A*-site atomic displacements exhibits a value of less than ~ 0.1 Å versus bulk SrRuO$_3$ (~ 0.12 Å), indicative of a suppressed tilt/rotation angle along the in-plane direction [**FIG. 5(h)**]. Under a compressively strained scenario, we have the following condition: $a_{pc} = b_{pc}$

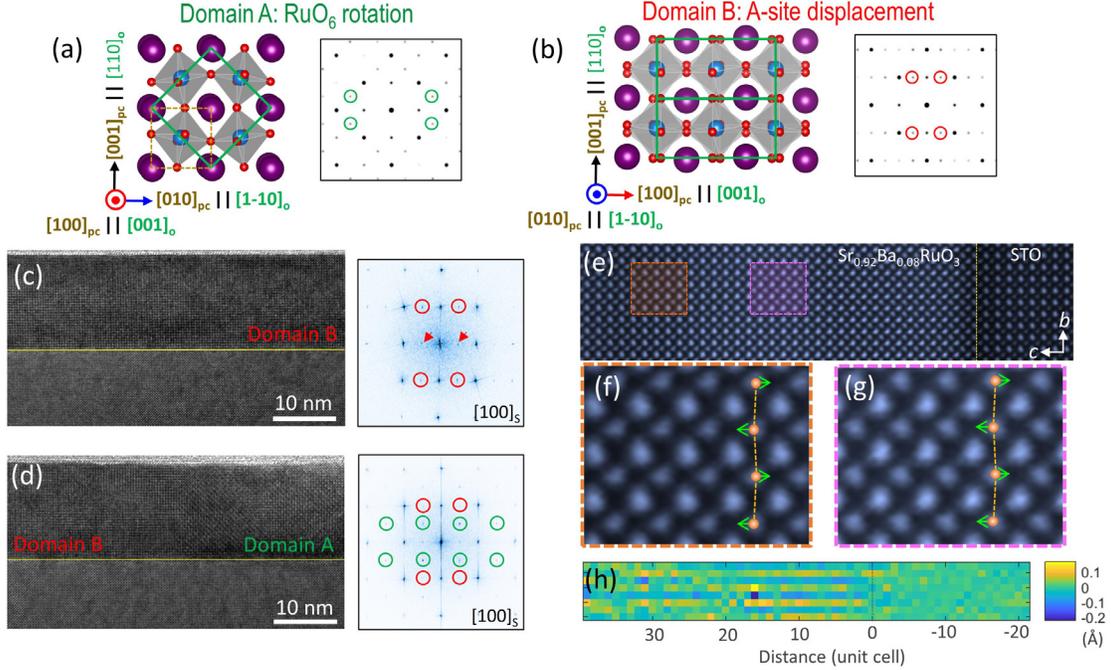

**FIG. 5.** Domain structure in Sr$_{0.92}$Ba$_{0.08}$RuO$_3$ film. (a-b) Projected crystal structures representing Domain A and B along with simulated electron diffraction patterns in the [100]$_{pc}$ (Domain A) and [010]$_{pc}$ (Domain B) directions. (c-d) High-resolution STEM images with electron beam incident along the [100]$_S$ direction, and corresponding Fast Fourier Transform (FFT) patterns with characteristic fractional spots marked in red (Domain B) and green (Domain A), respectively. The forbidden spots marked by the arrows, absent in the simulated electron diffractions were observed due to multiple scattering of the electron and sample. (e) HAADF image of Sr$_{0.92}$Ba$_{0.08}$RuO$_3$ film. The dotted yellow line marks the film-substrate interface. (f-g) Zoom-in images of the regions marked in panel (e) superimposed by the corresponding A-site cation displacement forming zig-zag pattern in the Sr/Ba-O plane. (h) A-site cation displacement as a function of distance from substrate.

< $c_{pc}$ (pseudocubic-cell elongation along out-of-plane), hence to have considerably smaller $a_{pc}$ and $b_{pc}$ compared to $c_{pc}$, the octahedron rotations magnitude along two in-plane axes ($a_{pc}$ and $b_{pc}$) would be reduced [48]. Considering the preceding discussion, it is clear that Sr$_{0.92}$Ba$_{0.08}$RuO$_3$ film holds an $a^-b^+c^-$ (domain B) rotation pattern (only a minor portion is domain A: $a^+b^-c^-$), but has diminished octahedra rotation magnitude. In other words, epitaxially strained $Sr_{0.92}Ba_{0.08}RuO_3$ film shows modification of crystal-cell symmetry (tetragonal) and octahedral rotation magnitude but the octahedra rotation pattern ($a^-b^+c^-/a^+b^-c^-$) remains intact.

To investigate the crystal structure of $Sr_{0.8}Ba_{0.2}RuO_3$ film, we have obtained HRTEM images across [100] and [210] STO-substrate directions, as revealed in **FIG. 6(a-d)**. The FFT results [see **FIG. 6(a-d)**] obtained from HRTEM images confirms the absence of fractional spots, indicating a cubic structure [58]. This is also evident in the annular-bright-field STEM image [**FIG. 6(e)**], where oxygen atoms forming a non-tilted octahedron environment could be realized, corroborating an $a^0b^0c^0$ perovskite structural symmetry. Furthermore, the half-integer X-ray reflections reflection results revealed in **FIG. 6(e)** suggests diminishing of characteristic peaks, supporting an $a^0b^0c^0$ tilt system. We note that for $Sr_{0.8}Ba_{0.2}RuO_3$ film, due to compressive epitaxial strain, the out-of-plane lattice expands, while the in-plane lattice parameter is locked with $SrTiO_3$ (i.e., $a_{pc} = b_{pc} < c_{pc}$), this leads to the pseudocubic-cell elongation along out-of-plane direction [Supplementary Note 1]. Overall, STEM structural analysis combined with half-integer X-ray measurements advocate that $Sr_{0.8}Ba_{0.2}RuO_3$ film holds an $a^0b^0c^0$ tilt-system, a non-rotational octahedra symmetry typically observed in cubic-type perovskites.

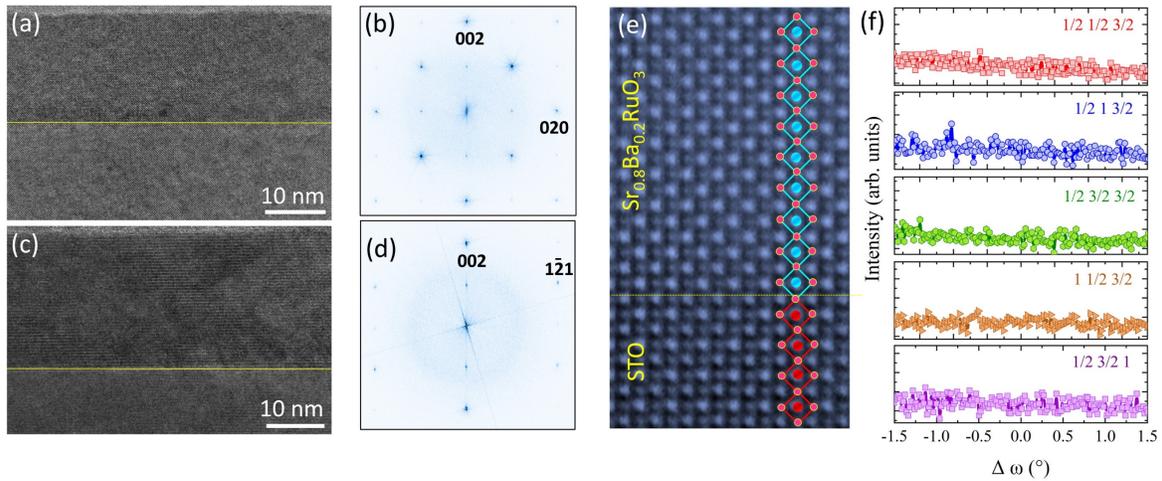

**FIG. 6.** Crystal structure of $Sr_{0.8}Ba_{0.2}RuO_3$ film determined via S/TEM. (a) High-resolution transmission electron microscope (HRTEM) image taken along the [100] direction, and (b) the corresponding FFT pattern. (c) HRTEM image with (d) Fast Fourier Transform (FFT) pattern taken along the [210] direction. The FFT patterns are indexed using the notation of the pseudo-cubic structure. (e) Highly-magnified annular-bright-field (ABF) STEM image taken along the [100] direction with projected structural model. The yellow line marks the $TiO_2$-(Sr/Ba)O interface. In panel (e) an octahedra without tilt could be observed. (f) Half-integer X-ray reflections for $Sr_{0.8}Ba_{0.2}RuO_3$ film.

### D. Magnetometry

In the following, we proceed to investigate the effects of structural transitions on the functional properties of $Sr_{1-x}Ba_xRuO_3$ thin films. **FIG. 7** shows temperature-dependent magnetization [$M(T)$] and magnetic hysteresis [$M(H)$] of $Sr_{1-x}Ba_xRuO_3$ thin films obtained: along the [100] and [010] in-plane directions and along the [001] out-of-plane direction. For $M(T)$, the samples were cooled down in 2000 Oe to 5 K, and in the second step, in presence of 100 Oe, the data is collected during warm-up. The orthorhombic SrRuO$_3$ film undergoes a sharp paramagnetic to ferromagnetic transition in $M(T)$ near ~145 K holding a perpendicular magnetic anisotropy (PMA) as in $M(T)$ and $M(H)$, where the easy axis resides in the [001] out-of-plane direction as in **FIG. 7(a-b)**. We also observed an in-plane magnetic anisotropy, which could be associated with different octahedra rotations along the in-plane directions where rotation is $a^-b^+c^-$. In contrast, the Sr$_{0.92}$Ba$_{0.08}$RuO$_3$ film reveals suppression of magnetic moments across two in-plane directions, while maintaining an out-of-plane easy magnetization axis [**FIG. 7(c-d)**]. Interestingly, under structural transition to purely tetragonal (without octahedral rotations: $a^0b^0c^0$) form in Sr$_{0.8}$Ba$_{0.2}$RuO$_3$ film, the system still exhibits robust ferromagnetism ($T_c \approx$ 145 K) with greatly enhanced PMA [**FIG. 7(e-f)**]. The configuration of the magnetic spins along two in-plane moments is highly suppressed, which leads to an isotropic in-plane magnetic response in Sr$_{0.8}$Ba$_{0.2}$RuO$_3$ film.

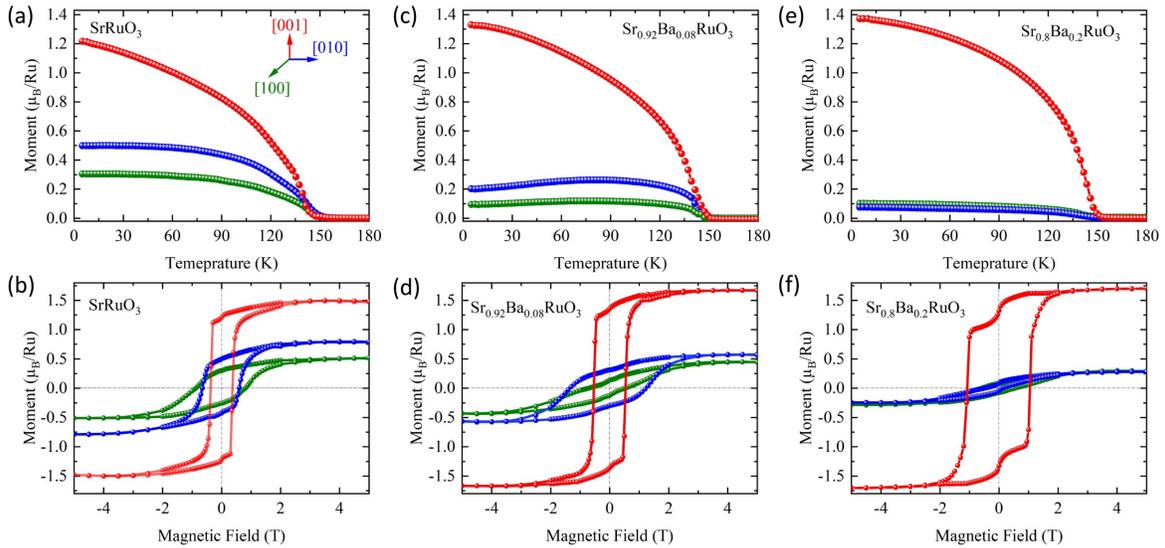

**FIG. 7.** Magnetization as a function of temperature [M(T)], and magnetic field [M(H)] for the family of $Sr_{1-x}Ba_xRuO_3$ ($0 \leq x \leq 0.2$) thin films. (a)-(b) $SrRuO_3$, (c)-(d) $Sr_{0.92}Ba_{0.08}RuO_3$, and (e)-(f) $Sr_{0.8}Ba_{0.2}RuO_3$.

### E. Electron Transport

To investigate the itinerancy of Sr$_{1-x}$Ba$_x$RuO$_3$ thin films, the temperature-dependent resistivity [$\rho(T)$] measurement is revealed in **FIG. 8(a)**. All films exhibit a metallic behavior of decreasing resistivity with kinks in resistivity occurring at temperatures coinciding with paramagnetic to ferromagnetic transition temperatures. The kink corresponds to $T_c$ and is clearly visible in the resistivity derivative ($d\rho/dT$) [inset **FIG. 8(a)**], which arises from magnetic scattering and is related to a linear-$\rho(T)$ modification to a $\rho(T)$-$T^2$ behavior [59,60]. Additionally, we have measured the magnetoresistance: MR $=[R(0) - R(H)]/R(0)$, where $R(H)$, and $R(0)$ are the resistances with magnetic field, and without magnetic field, respectively [**FIG. 8(b-d)**]. For MR, the external magnetic field was applied along the film normal. We note that in the paramagnetic phase, the MR exhibits a parabolic shape, while a negative, non-parabolic MR appears in FM phase [20,61–67].

In this case, the presence of hysteretic MR at the lowest temperature of 5 K [**FIG. 8(b-d)**] confirms the existence of robust ferromagnetic ordering in SrRuO$_3$, Sr$_{0.92}$Ba$_{0.08}$RuO$_3$ and Sr$_{0.8}$Ba$_{0.2}$RuO$_3$ thin films. It is noted that the MR magnetic hysteresis endures up to ~ 100 K, whereas above 100 K, a non-parabolic MR persists up to ~ 145 K. These observations approve that the systems under study hold a clear ferromagnetic order up to ~ 145 K.

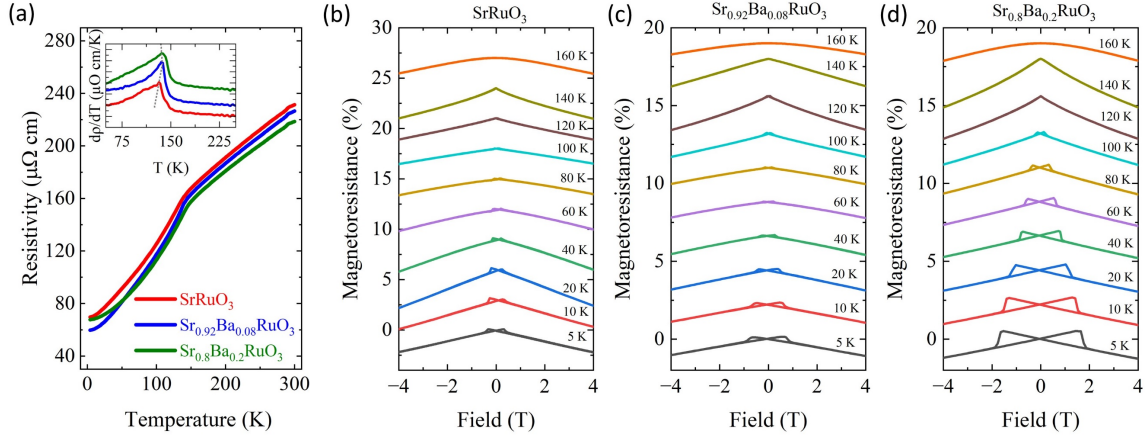

**FIG. 8.** (a) Sr$_{1-x}$Ba$_x$RuO$_3$ (0 ≤ $x$ ≤ 0.2) thin films, temperature-dependent resistivity. The inset in panel (a) is temperature-dependent resistivity derivative ($d\rho/dT$) plotted against temperature ($T$). Magnetoresistance (MR) of (b) SrRuO$_3$, (c) Sr$_{0.92}$Ba$_{0.08}$RuO$_3$, and (d) Sr$_{0.8}$Ba$_{0.2}$RuO$_3$ thin films measured at different temperatures with magnetic field applied along out-of-plane direction. The MR curves are offset to avoid overlap.

## F. Anomalous Hall

The Anomalous Hall effect (AHE) is investigated to further describe the films intrinsic magnetization behaviors. The total hall or transverse resistance is specified as $R_{xy} = R_0 H + R_A M$. Here, the first term ($R_0 H$) signifies the ordinary Hall component arising from Lorentz force contribution with $R_0$ being the ordinary Hall coefficient and $H$ the magnetic field. The second term ($\rho_{AHE} = R_A M$) indicates the anomalous contribution, where $R_A$ is the anomalous Hall coefficient, and $M$ the magnetization. For clarification of the anomalous component, we subtracted the ordinary Hall term by linear fitting, which is obtained from the total hall resistance versus the applied field curve in the high field regime. The anomalous contribution measured at different temperatures for a series of Sr$_{1-x}$Ba$_x$RuO$_3$ ($x$ = 0, 0.08 and 0.2) thin films is presented in **FIG. 9**. The reversed hysteretic curve at the lowest temperatures (5 K) signifies strong FM order encompassing a negative $R_A$ [**FIG. 9(a)**]. The SrRuO$_3$ ($x$ = 0) film displays an inverted magnetic hysteresis loop to 100 K, while at 120 K, a sign reversal is observed [**FIG. 9(a)**], which indicates $R_A$ sign reversal from negative to positive. Comparable Hall structure with pronounced inverted squarish shape hysteresis at low temperatures could be interpreted as being the result of strong ferromagnetism in Sr$_{0.92}$Ba$_{0.08}$RuO$_3$ [**FIG. 9(b)**] and Sr$_{0.8}$Ba$_{0.2}$RuO$_3$ [**FIG. 9(c)**] films.

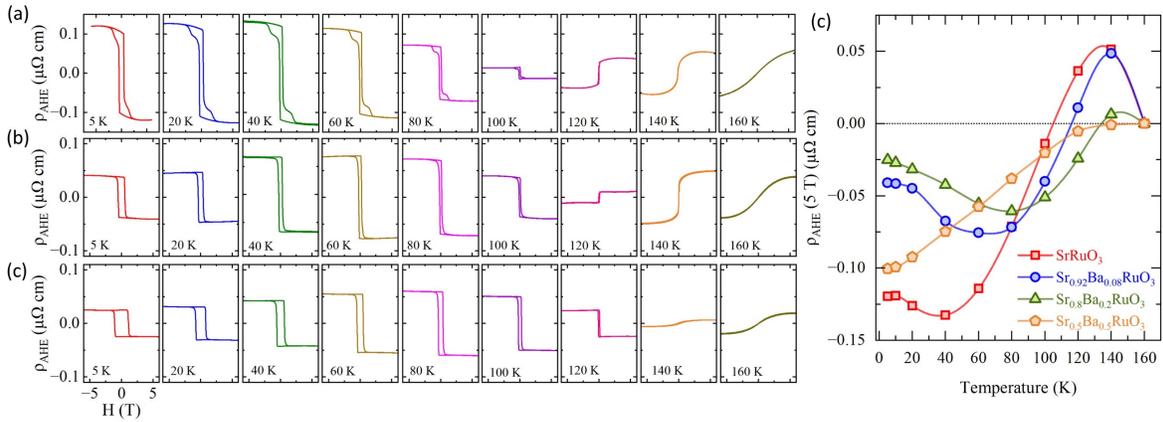

**FIG. 9.** Anomalous Hall resistivity ($\rho_{AHE}$) of (a) SrRuO$_3$, (b) Sr$_{0.92}$Ba$_{0.08}$RuO$_3$, and (c) Sr$_{0.8}$Ba$_{0.2}$RuO$_3$. (d) (b) Temperature dependence of anomalous Hall resistivity ($\rho_{AHE}$) measured at the magnetic-field value of 5 T for series of Sr$_{1-x}$Ba$_x$RuO$_3$ ($0 \leq x \leq 0.5$) thin films.

Interestingly, the introduction of Ba pushes [**FIG. 9(a-c)**] the hysteretic flip to higher temperatures (i.e., temperature at which $R_A$ changes sign from positive to negative). Finally, for Sr$_{0.5}$Ba$_{0.5}$RuO$_3$ film, the $R_A$ remains negative throughout measured temperature range [**FIG. 9(d)**]. The change of Hall-sign with increasing Ba-concentration is likely associated with two structural effects that can change the Fermi surface morphology; (1) RuO$_6$ octahedral rotations reduction, and (2) Ru-O bond length variation. Tian et al. [68] reported that SrRuO$_3$ with or without octahedral rotations brings similar behavior of a sign-change from negative to positive in Hall conductivity under the changes from compressive to tensile epitaxial strain. On other hand, the larger Ba-concentration (especially as $x \geq 0.5$) results in pseudocubic-cell stretching (Tetragonal-type distortion: $c_{pc}/a_{pc} \gg 1$). This effect leads to further crystal field splitting ($\Delta_T$) between Ru $d$-orbitals, which in turn changes the Ru orbital energies, and electron occupancy, leading to sign change of anomalous Hall conductivity [13,22,68,69].

## G. Ferromagnetism versus Ba-substitution

Finally, we discuss the evolution of ferromagnetism and $T_c$ in a series of Sr$_{1-x}$Ba$_x$RuO$_3$ as presented in **FIG. 10(a)**. The SrRuO$_3$ film possesses a $T_c$ of 145 K, while Sr$_{0.92}$Ba$_{0.08}$RuO$_3$ film demonstrates a slightly enhanced $T_c$ of 149 K. Remarkably, the Sr$_{0.8}$Ba$_{0.2}$RuO$_3$ film with fully suppressed octahedral rotations ($a^0b^0c^0$) and flat Ru-O-Ru bond angles has a nearly identical $T_c \approx 145$ K as compare to the orthorhombic variant. However, the Sr$_{0.5}$Ba$_{0.5}$RuO$_3$ film shows lowering of Curie temperature to $T_c \approx 100$ K, which reduced further ($T_c \approx 100$ K) for Sr$_{0.5}$Ba$_{0.5}$RuO$_3$ film [Supplementary Note 2]. Moreover, the Sr$_{0.5}$Ba$_{0.5}$RuO$_3$ films displays significant suppression of magnetic moment. This suggests, even though the structure remains in a higher-symmetry regime (no octahedron rotations), the larger amount of Ba-cation implantation introduces some subtle changes in electronic and magnetic structure, which condenses the ferromagnetic order. It is worth to note that in the Sr$_{1-x}$Ba$_x$RuO$_3$ bulk series, the orthorhombic-type *Pbnm* phase shows a maximized $T_c = 160$ K [37,38]. However, the introduction of Ba-cation systematically reduces the octahedral distortion, leading to a the cubic-type *Pm-3m* phase forms at $x = 0.3$-$0.4$. In bulk-series, the increased Ba-concentration elongates the Ru-O bond, and saturates the $T_c$ for final member (BaRuO$_3$). The trend of $T_c$ for the thin-films results presented here is in good agreement with the bulk-series [37,38].

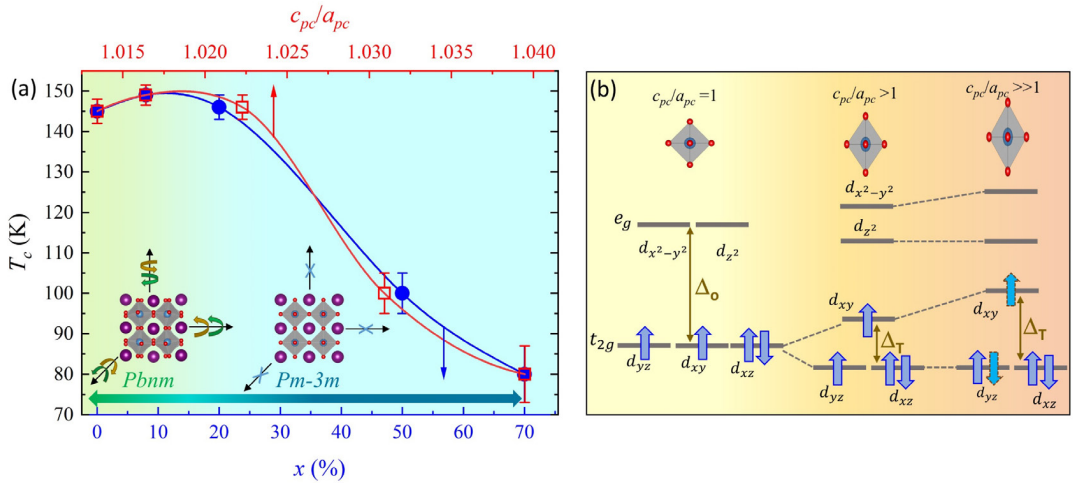

**FIG. 10.** (a) Phase diagram of $T_c$ in series of Sr$_{1-x}$Ba$_x$RuO$_3$ thin films as function of Ba-concentration ($x$), and ratio of out-of-plane ($c_{pc}$) and in-plane ($a_{pc}$) pseudocubic lattice constants. (b) Schematic illustration of Ru 4-$d$ orbitals. The RuO$_6$ octahedra undergoes tetragonal crystal field splitting ($\Delta_T$) due to $c_{pc}/a_{pc} > 1$.

Table 1. Pseudocubic lattice parameter, crystal symmetry, octahedra rotation pattern, Curie Temperature ($T_c$), and magnetic moment of Sr$_{1-x}$Ba$_x$RuO$_3$ films.

| $x$ in Sr$_{1-x}$Ba$_x$RuO$_3$ | Pseudocubic Lattice Parameter (Å) | Crystal Symmetry | Octahedra Rotation Pattern | $T_c$ (K) | Magnetic Moment ($\mu_B$/Ru) |
|---|---|---|---|---|---|
| 0 | 3.957 | Orthorhombic | $a^-b^+c^-$ | 145±3 | 1.51 |
| 0.08 | 3.969 | Tetragonal | $a^-b^+c^-/a^+b^-c$ | 149±4 | 1.67 |
| 0.2 | 3.992 | Tetragonal | $a^0b^0c^0$ | 146±3 | 1.68 |
| 0.5 | 4.025 | Tetragonal | $a^0b^0c^0$ | 100±5 | 0.45 |
| 0.7 | 4.059 | Tetragonal | $a^0b^0c^0$ | 80±7 | 0.17 |

**H.    Discussion**

We have systematically shown that epitaxial-strain provides an effective platform to stabilize $Sr_{1-x}Ba_xRuO_3$ films. Investigation of a series of isovalent substitutional doping revealed that this system is sensitive to structural tunings that can be used to manipulate electro-magnetic properties. Low Ba-concentration ($x \leq 0.2$) benefits an effective structural transition from orthorhombic ($a^-b^+c^-$) to a purely tetragonal symmetry ($a^0b^0c^0$) by diminishing octahedral rotational distortion. Simultaneously, the initial Ba-substitution ($x \leq 0.2$) to $SrRuO_3$ also enhances the ferromagnetism reflected by a slight increase of $T_c$ which is maximized around $x = 0.2$ [see **Table 1**]. The physics associated with such a structure-property relationship can be understood based on the octahedron rotational distortion dependent FM model [40,70–72]. In this picture, the reduction of the orthorhombic distortion favors the enhancement of the electron itinerancy due to the increase of bandwidth and density of states at Fermi energy, which leads to enhancement of itinerant-type ferromagnetism as observed for initial Ba-substitution ($x \leq 0.2$).

At the same time, under compressive strain constraint by $SrTiO_3$, the in-plane lattice parameters of all films ($a_{pc} = b_{pc}$) are essentially the same as that of $SrTiO_3$ substrate. Consequently, the compressively strained $SrRuO_3$ shows a slight increase of the out-of-plane parameter ($c_{pc}$), that pushes toward a tetragonal-type pseudocubic unit-cell ($c_{pc}/a_{pc} > 1$). Such pseudocubic-cell stretching ($c_{pc}/a_{pc} > 1$) noticeably amplifies with increasing Ba-substitution [see **FIG. 1(c)**, and **FIG. S1**], leading to an enhanced structural anisotropy. The enhanced tetragonality influences the $4d$-$t_{2g}$ orbital arrangement and occupancy. In bulk $SrRuO_3$ ($c_{pc}/a_{pc} = 1$), under an octahedral field ($\Delta_c$), the Ru-$4d$ state holds a low-spin configuration ($4d^4: t_{2g}^{3\uparrow,1\downarrow} e_g^0; S = 1$), whereas the $t_{2g}$ ($d_{xy}, d_{yz}, d_{xz}$) states are nearly degenerate since the tetragonal-field splitting ($\Delta_T$) is very small [**FIG. 10(b)**]. In the thin film form such as that described in this work, the compressively-strained $SrRuO_3$ or smaller Ba-concertation ($x \leq 0.2$) films under pseudocubic-cell elongation ($c_{pc}/a_{pc}>1$) exhibit a splitting in $d_{yz}$ and $d_{xz}$ from the $d_{xy}$ states due to a slight increase of $\Delta_T$. Whereas due to the antibonding nature, the $d_{xy}$ are higher in energy than $d_{yz}$ and $d_{xz}$ [**FIG. 10(b)**]. The larger Ba-substitution ($x \geq 0.5$) can further amplify the $\Delta_T$ splitting due to a large enhancement of pseudocubic-cell lengthening along out-of-plane ($c_{pc}/a_{pc} \gg 1$). As a result, a larger $\Delta_T$ split could partially transform the $d_{xy}$ states to $d_{yz}, d_{xz}$ and leads to even lower spin states that would result in the reduction of magnetic moment and FM order as seen for $x = 0.5$. Hence, the enhanced tetragonality with larger Ba-substitution can diminish the ferromagnetism via orbitals redistribution. Such competing effects from structure to magnetism results in a non-monotonic change of Curie temperature with Ba-substitution. As the octahedral rotations disappear at $x \geq 0.2$, the effect of pseudocubic-cell stretching ($c_{pc}/a_{pc} \gg 1$) dominates with $x$ and steadily suppresses the ferromagnetism of the system. In addition, the enhanced structural anisotropy also results in the enhanced perpendicular magnetic anisotropy with Ba-substitution as observed from our experiments.

## IV. CONCLUSION

In conclusion, we fabricated the $Sr_{1-x}Ba_xRuO_3$ ($0 \leq x \leq 0.7$) thin films. The strain-stabilized Ba substitution transmutes the lattice symmetry from bulk-like orthorhombically distorted ($x = 0$) to tetragonal phase ($x = 0.2$) without $RuO_6$ rotations. The epitaxially stabilized tetragonal phase without $RuO_6$ rotations is ferromagnetically ($T_c \sim 145$ K) ordered with a strong perpendicular magnetic anisotropy. Though, increased Ba-substitution ($x \geq 0.5$) significantly condenses the ferromagnetism via orbital reconstructions. The investigations provide an essential understanding of ruthenates ferromagnetism, indicating sensitivity to *A*-site in determining the crystal structure as well as electro-magnetic properties of ruthenates.

## ACKNOWLEDGMENTS

This work is primarily supported by the US Department of Energy (DOE) under Grant No. DOE DE-SC0002136. The electron microscopy work done at Brookhaven National Laboratory (BNL) was sponsored by the US DOE-BES, Materials Sciences and Engineering Division, under Contract No. DE-SC0012704. Use of the BNL Center for Functional Nanomaterials supported by the BES Office of User Science Facilities for TEM sample preparation is also acknowledged. Work at Oak Ridge National Laboratory was supported by the US Department of Energy (DOE), Office of Basic Energy Sciences (BES), Materials Sciences and Engineering Division. The work at Los Alamos National Laboratory was supported by the NNSA's Laboratory Directed Research and Development Program, and was performed, in part, at the CINT, an Office of Science User Facility operated for the U.S. Department of Energy Office of Science through the Los Alamos National Laboratory. Los Alamos National Laboratory is operated by Triad National Security, LLC, for the National Nuclear Security Administration of U.S. Department of Energy (Contract No. 89233218CNA000001).

## DATA AVAILABILITY

The data used in the study are available in the manuscript and supplementary data. The data that support the findings of this study are available from the corresponding author upon reasonable request.

## AUTHORS CONTRIBUTION

Z.A., M.S., and J.Z. designed the research. Z.A. grew the samples and performed transport/magnetization measurements. Z.A. and M.S. performed the X-ray diffraction. Z.A. and R.N. prepared thin-film targets. Z.W. and Y.Z. carried the STEM experiments. A.R.M. and T.Z.W. did half-angle X-ray diffraction. Z.A. wrote the manuscript with input from all authors.

[15] A. Vailionis, W. Siemons, and G. Koster, *Room Temperature Epitaxial Stabilization of a Tetragonal Phase in ARuO$_3$ (A=Ca and Sr) Thin Films*, Appl. Phys. Lett. **93**, 051909 (2008).

[16] S. S. Hong, M. Gu, M. Verma, V. Harbola, B. Y. Wang, D. Lu, A. Vailionis, Y. Hikita, R. Pentcheva, J. M. Rondinelli, and H. Y. Hwang, *Extreme Tensile Strain States in La$_{0.7}$Ca$_{0.3}$MnO$_3$ Membranes*, Science **368**, (2020).

[17] D. Kan, R. Aso, R. Sato, M. Haruta, H. Kurata, and Y. Shimakawa, *Tuning Magnetic Anisotropy by Interfacially Engineering the Oxygen Coordination Environment in a Transition Metal Oxide*, Nat. Mater. **15**, 432 (2016).

[18] Z. Liao, M. Huijben, Z. Zhong, N. Gauquelin, S. Macke, R. J. Green, S. Van Aert, J. Verbeeck, G. Van Tendeloo, K. Held, G. A. Sawatzky, G. Koster, and G. Rijnders, *Controlled Lateral Anisotropy in Correlated Manganite Heterostructures by Interface-Engineered Oxygen Octahedral Coupling*, Nat. Mater. **15**, 425 (2016).

[19] M. Meng, Z. Wang, A. Fathima, S. Ghosh, M. Saghayezhian, J. Taylor, R. Jin, Y. Zhu, S. T. Pantelides, J. Zhang, E. W. Plummer, and H. Guo, *Interface-Induced Magnetic Polar Metal Phase in Complex Oxides*, Nat. Commun. **10**, 5248 (2019).

[20] Z. Ali, Z. Wang, A. O'Hara, M. Saghayezhian, D. Shin, Y. Zhu, S. T. Pantelides, and J. Zhang, *Origin of Insulating and Non-Ferromagnetic SrRuO$_3$ Monolayers*, Phys. Rev. B **105**, 054429 (2022).

[21] A. Herklotz, A. T. Wong, T. Meyer, M. D. Biegalski, H. N. Lee, and T. Z. Ward, *Controlling Octahedral Rotations in a Perovskite via Strain Doping*, Sci. Rep. **6**, 1 (2016).

[22] E. Skoropata, A. R. Mazza, A. Herklotz, J. M. Ok, G. Eres, M. Brahlek, T. R. Charlton, H. N. Lee, and T. Z. Ward, *Post-Synthesis Control of Berry Phase Driven Magnetotransport in SrRuO$_3$ Films*, Phys. Rev. B **103**, 1 (2021).

[23] C. Wang, C. Chang, A. Herklotz, C. Chen, F. Ganss, U. Kentsch, D. Chen, X. Gao, Y. Zeng, O. Hellwig, M. Helm, S. Gemming, Y. Chu, and S. Zhou, *Topological Hall Effect in Single Thick SrRuO$_3$ Layers Induced by Defect Engineering*, Adv. Electron. Mater. **6**, 2000184 (2020).

[24] S. J. May, C. R. Smith, J. W. Kim, E. Karapetrova, A. Bhattacharya, and P. J. Ryan, *Control of Octahedral Rotations in (LaNiO$_3$)$^n$/(SrMnO$_3$)$^m$ Superlattices*, Phys. Rev. B **83**, 2 (2011).

[25] H. Guo, Z. Wang, S. Dong, S. Ghosh, M. Saghayezhian, L. Chen, Y. Weng, A. Herklotz, T. Z. Ward, R. Jin, S. T. Pantelides, Y. Zhu, J. Zhang, and E. W. Plummer, *Interface-Induced Multiferroism by Design in Complex Oxide Superlattices*, Proc. Natl. Acad. Sci. U. S. A. **114**, E5062 (2017).

[26] C. Domínguez, A. B. Georgescu, B. Mundet, Y. Zhang, J. Fowlie, A. Mercy, A. Waelchli, S. Catalano, D. T. L. Alexander, P. Ghosez, A. Georges, A. J. Millis, M. Gibert, and J. M. Triscone, *Length Scales of Interfacial Coupling between Metal and Insulator Phases in Oxides*, Nat. Mater. **19**, 1182 (2020).

[27] B. Chen, N. Gauquelin, R. J. Green, J. H. Lee, C. Piamonteze, M. Spreitzer, D. Jannis, J. Verbeeck, M. Bibes, M. Huijben, G. Rijnders, and G. Koster, *Spatially Controlled Octahedral Rotations and Metal-Insulator Transitions in Nickelate Superlattices*, Nano Lett. **21**, 1295 (2021).

[28] K. Takiguchi, Y. K. Wakabayashi, H. Irie, Y. Krockenberger, T. Otsuka, H. Sawada, S. A. Nikolaev, H. Das, M. Tanaka, Y. Taniyasu, and H. Yamamoto, *Quantum Transport Evidence of Weyl Fermions in an Epitaxial Ferromagnetic Oxide*, Nat. Commun. **11**, 4969 (2020).

[29] M. Verma, B. Geisler, and R. Pentcheva, *Effect of Confinement and Octahedral Rotations on the Electronic, Magnetic, and Thermoelectric Properties of Correlated SrXO$_3$/SrTiO$_3$(001) Superlattices (X= V, Cr, or Mn)*, Phys. Rev. B **100**, 165126 (2019).